\documentclass[]{spie}  

\usepackage{amsmath,amsfonts,amssymb}
\usepackage{graphicx}
\usepackage[colorlinks=true, allcolors=blue]{hyperref}

\usepackage{orcidlink}
\newcommand{\orcid}[1]{\href{https://orcid.org/#1}{\textcolor[HTML]{A6CE39}{\aiOrcid}}}

\title{Cosmology Large Angular Scale Surveyor (CLASS): Pointing Stability and Beam Measurements at 90, 150, and 220 GHz}

\author[a]{Rahul~Datta~\orcidlink{0000-0003-3853-8757}}
\author[a]{Michael~K.~Brewer}
\author[a]{Jullianna~D.~Couto~\orcidlink{0000-0002-0552-3754}}
\author[a]{Joseph~R.~Eimer~\orcidlink{0000-0001-6976-180X}}
\author[a]{Yunyang~Li~\orcidlink{0000-0002-4820-1122}}
\author[b]{Zhilei~Xu~\orcidlink{0000-0001-5112-2567}}

\author[a]{John~W.~Appel~\orcidlink{0000-0002-8412-630X}}
\author[c]{Ricardo~Bustos~\orcidlink{0000-0001-8468-9391}}
\author[d]{David~T.~Chuss~\orcidlink{0000-0003-0016-0533}}
\author[a]{Joseph~Cleary}
\author[e]{Sumit~Dahal~\orcidlink{0000-0002-1708-5464}}
\author[e]{Thomas~Essinger-Hileman~\orcidlink{0000-0002-4782-3851}}
\author[f]{Jeffrey~Iuliano~\orcidlink{0000-0001-7466-0317}}
\author[a]{Tobias~A.~Marriage~\orcidlink{0000-0003-4496-6520}}
\author[a]{\mbox{Carolina N\'{u}\~{n}ez~\orcidlink{0000-0002-5247-2523}}}
\author[g]{Matthew~A.~Petroff~\orcidlink{0000-0002-4436-4215}}
\author[e]{Karwan~Rostem~\orcidlink{0000-0003-4189-0700}}
\author[h]{Duncan~J.~Watts~\orcidlink{0000-0002-5437-6121}}
\author[e]{Edward~J.~Wollack~\orcidlink{0000-0002-7567-4451}}

\affil[a]{The William H. Miller III Department of Physics and Astronomy, Johns Hopkins University, 3701 San Martin Drive, Baltimore, MD
21218, USA}
\affil[b]{MIT Kavli Institute, Massachusetts Institute of Technology, 77 Massachusetts Avenue, Cambridge, MA 02139, USA}

\affil[c]{Departamento de Ingenier\'{i}a El\'{e}ctrica, Universidad Cat\'{o}lica de la Sant\'{i}sima Concepci\'{o}n, Alonso de Ribera 2850, Concepci\'{o}n, Chile}

\affil[d]{Department of Physics, Villanova University, 800 Lancaster Avenue, Villanova, PA 19085, USA}

\affil[e]{Goddard Space Flight Center, 8800 Greenbelt Road, Greenbelt, MD 20771, US}

\affil[f]{Department of Physics and Astronomy, University of Pennsylvania, 209 South 33rd Street, Philadelphia, PA 19104, USA}

\affil[g]{Center for Astrophysics, Harvard \& Smithsonian, 60 Garden Street, Cambridge, MA 02138, USA}

\affil[h]{Institute of Theoretical Astrophysics, University of Oslo, P.O. Box 1029 Blindern, N-0315 Oslo, Norway}

\authorinfo{Send correspondence to R. Datta, E-mail: rdatta2@jhu.edu}

\begin{document} 
\maketitle

\begin{abstract}
The Cosmology Large Angular Scale Surveyor (CLASS) telescope array surveys 75$\%$ of the sky from the Atacama desert in Chile at frequency bands centered near 40, 90, 150, and 220 GHz. CLASS measures the largest-angular-scale ($\theta\gtrsim1^\circ$) CMB polarization with the aim of constraining the tensor-to-scalar ratio, $r$, measuring the optical depth to reionization, $\tau$, to near the cosmic variance limit, and more. The CLASS Q-band (40 GHz), W-band (90 GHz), and dichroic high frequency (150/220 GHz) telescopes have been observing since June 2016, May 2018, and September 2019, respectively. On-sky optical characterization of the 40~GHz instrument has been published. Here, we present preliminary on-sky measurements of the beams at 90, 150, and 220 GHz, and pointing stability of the 90 and 150/220 GHz telescopes. The average 90, 150, and 220 GHz beams measured from dedicated observations of Jupiter have full width at half maximum (FWHM) of $0.615\pm0.019^{\circ}$, $0.378\pm0.005^{\circ}$, and $0.266\pm0.008^{\circ}$, respectively. Telescope pointing variations are within a few $\%$ of the beam FWHM.  
\end{abstract}


\section{INTRODUCTION}
\label{sec:intro}  

The Cosmology Large Angular Scale Surveyor (CLASS) is designed to measure the polarization of the CMB at large angular scales~\cite{essi14,harr16} enabled by rapid front-end polarization modulation~\cite{harr18} and a scanning strategy allowing adequate cross-linking~\cite{mill16}. Observing over $75\%$ of the sky in multiple frequency bands centered near 40, 90, 150, and 220~GHz from a high altitude site in the Atacama desert in Chile, CLASS will be sensitive to both the recombination and reionization~\cite{kami16} signatures in the B-mode signal associated with primordial gravitational waves from inflation. The large angular scale E-mode measurement will improve constraints on the optical depth to reionization~\cite{2018_tau_Duncan}, which will break degeneracies in cosmological parameters, specifically on measurements from small-angular-scale CMB polarization of the sum of neutrino masses~\cite{Neutrino_Allison}. The 40~GHz telescope~\cite{2014_40GHz_Detector_John,appe19} has been observing since June 2016, while the 90~GHz telescope achieved first light in May 2018~\cite{daha18} and is currently being upgraded. A third telescope housing a dichroic receiver sensitive to both 150 and 220~GHz frequency bands~\cite{daha19} (henceforth referred to as HF) began observations in September 2019. Another telescope at 90~GHz is planned.

\begin{figure*}
    \centering
    \includegraphics[width=0.9\linewidth]{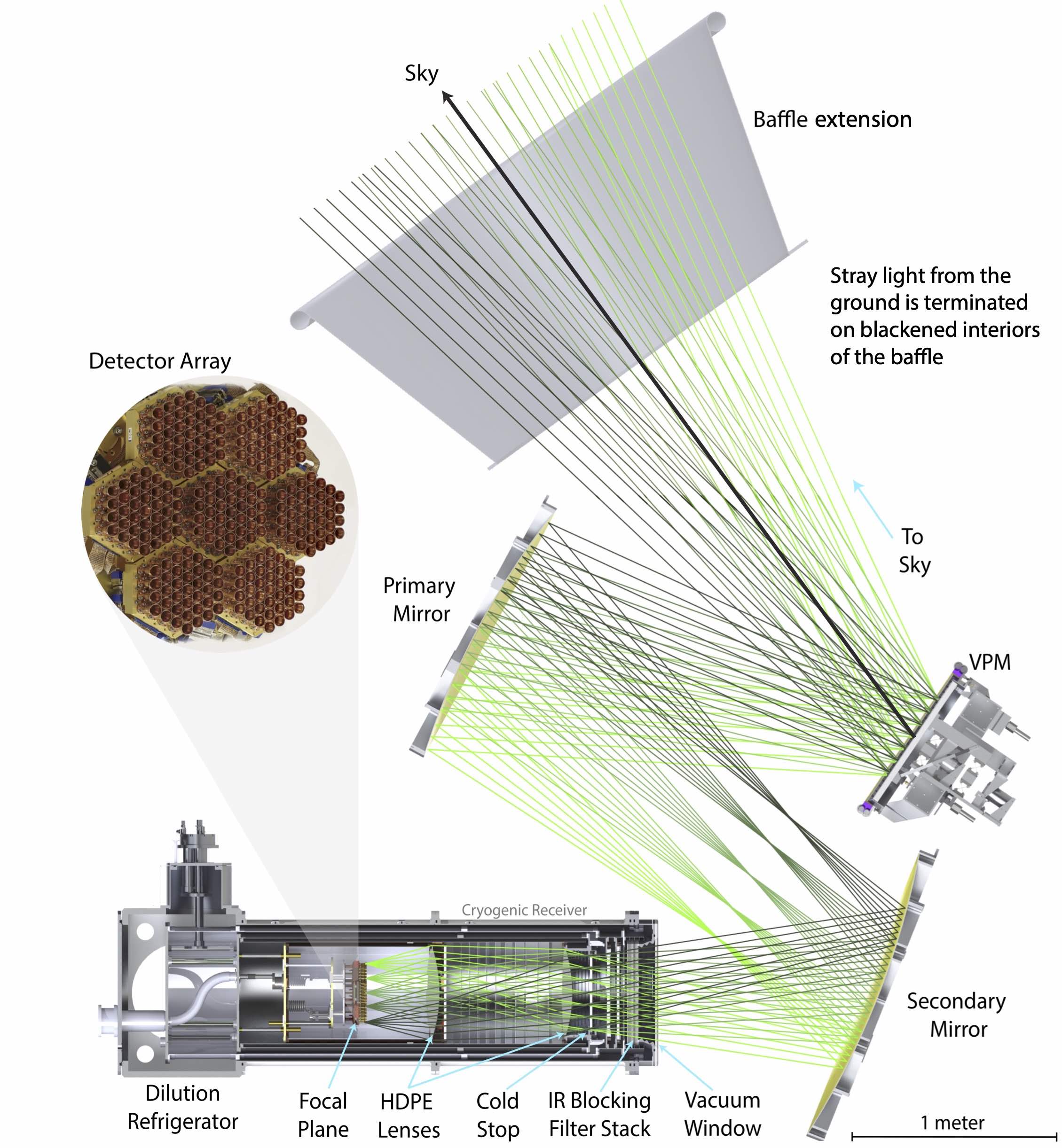}
    \caption{Ray-trace schematic of the 90~GHz telescope showing the major optical components. Color-coded rays from a selection of feedhorns illustrate the marginal light path and field-of-view of the telescope. The VPM is the first optical element in the path of the incoming light through a co-moving baffle. The co-moving baffle comprises a ``cage" structure (see Figure~\ref{fig:bs_rotation}), which houses the fore-optics and the receiver, and a  ``baffle extension." The solid black line shows the boresight pointing. Stray light is terminated on blackened interiors of the baffle. The mirrors image the cold stop onto the central part of the VPM. High density polyethylene (HDPE) lenses focus the light onto  feedhorn-coupled dual-polarization detectors. Seven detector modules, each comprising 37 pixels, constitute the focal plane array. The HF telescope optical design is conceptually similar; however, it employs silicon instead of HDPE lenses. The HF focal plane consists of three detector modules, each comprising 85 pixels.  } 
    \label{fig:ray_schematic}
\end{figure*}

The optical characterization of the CLASS 40~GHz telescope during the first two years of its operation (``Era~1'') was described by Xu et al~\cite{2020ApJ...891..134X}. In these proceedings, we present preliminary measurements of the 90~GHz and HF telescope pointing stability over several months during the 2020--2021 observing season. We also present preliminary on-sky measurements of the instrument beam in intensity at 90, 150 and 220~GHz made from dedicated planet observations. Characterization of far sidelobes, constraints on instrumental temperature-to-polarization leakage, polarization angle calibration, and beam window functions will be discussed in a future paper.

This paper is organized as follows. The CLASS instrument and survey scan strategy, including telescope configuration, observing modes, and data used in this analysis are described in Section~\ref{sec:instrument&obs}. Preliminary pointing stability of the 90~GHz and HF telescopes is presented in Section~\ref{sec:pointing}. In Section~\ref{sec:beams}, preliminary beam measurements are presented. Finally, we summarize in Section~\ref{sec:summary}.

\section{INSTRUMENT \& OBSERVATIONS}
\label{sec:instrument&obs}

A schematic of the 90~GHz CLASS telescope is shown in Fig.~\ref{fig:ray_schematic} with corresponding ray-trace and main components rendered. The optical design is similar to the design of the 40~GHz telescope \cite{eime12}. The first optical element in the path of the sky signal is the variable-delay polarization modulator (VPM) \cite{eime12, chus12, harr18}, which modulates the polarized signal at 10~Hz before potential instrument-induced polarization signals. This front-end modulation to frequencies higher than atmospheric and instrumental drifts enables recovery of the largest angular scale modes while suppressing temperature-to-polarization leakage from atmospheric signals, which is key to achieving the science goals of CLASS. The fore-optics comprising the primary and secondary mirrors produce an image of the cold stop near the central portion of the VPM, which is 60~cm in diameter. An ultra-high molecular weight polyethylene vacuum window on the cryogenic receiver followed by a stack of infrared (IR)-blocking filters~\cite{essi14, 2018Cryostat_Jeff} allow microwave light to enter the receiver while keeping out infrared radiation. The 60~cm VPM is significantly under-illuminated. This minimizes edge effects that could produce unwanted sidelobes. Two high-density polyethylene (HDPE) lenses re-image the light from each field onto the focal plane with an effective $f$/$\sim1.8$. The HF telescope optics and receiver design is conceptually similar; however, it employs silicon instead of HDPE lenses. The HF infrared filtering scheme is also different (see Iuliano et al.~\cite{2018Cryostat_Jeff}).

\begin{figure*}[htb]
    \centering
    \includegraphics[width=1\linewidth]{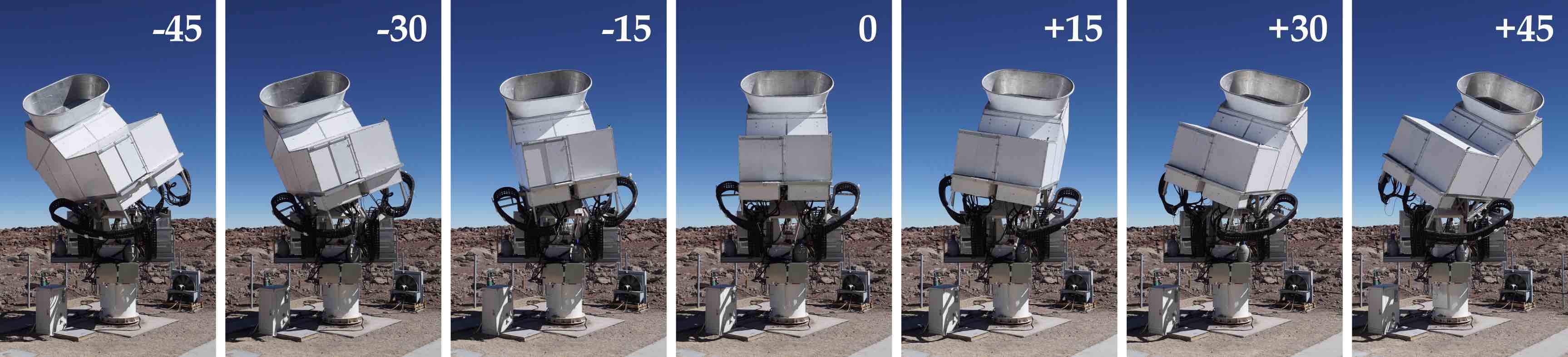}
    \caption{Photos of the telescope mount showing the seven boresight rotation angles. This rotation keeps the telescope boresight pointing unchanged while rotating the detector antenna orientation on the sky in steps of $15^{\circ}$ within a $90^{\circ}$ range, thus enabling measurement of the polarization signal projected onto different orientations.}
    \label{fig:bs_rotation}
\end{figure*}

The focal planes of the 90~GHz and HF telescopes consist of seven and three detector modules, respectively. Each 90~GHz module has 37 smooth-walled feedhorns~\cite{zeng2010, daha18} that couple the incoming light to microfabricated antenna probes on a 100~mm planar detector wafer~\cite{Rostem:2016}. Each HF module has 85 smooth-walled feedhorns~\cite{daha19}. The feeds are hexagonal close packed. The 90~GHz feeds are individually assembled whereas the HF feeds are fabricated in array format~\cite{doi:10.1063/5.0049526}. The 90~GHz telescope's field-of-view (FOV) is  approximately $23^\circ$ in diameter, while the FOV of the HF telescope is approximately $18^\circ$ in diameter. Each pixel is sensitive to both orthogonal states of linear polarization enabled by an orthomode transducer (OMT) on the detector wafer. Each pixel has two transition-edge sensor (TES) bolometers for measuring the power in both polarization states. The focal plane is cooled to $\sim50~mK$ by a dilution refrigerator~\cite{2018Cryostat_Jeff}. The on-sky receiver performance at all four frequencies is presented in Dahal et al.~\cite{2022ApJ...926...33D}.

A three-axis mount, shown in Figure~\ref{fig:bs_rotation}, points the 90 and 40~GHz telescopes. A second identical mount supports the HF telescope. The two mounts can rotate in azimuth, elevation, and boresight independent of each other. During CMB observations, the mounts nominally rotate 720$^{\circ}$ in azimuth at a fixed elevation of 45$^{\circ}$, before reversing and scanning in the opposite direction, thus scanning the sky in large circles. During the majority of CMB observations, the telescope scanned in azimuth at 2$^{\circ}$/s. During the day, the telescope boresight is restricted to point at least 20$^{\circ}$ away from the Sun, increasing the frequency of scan direction reversals and reducing the azimuthal range from the nominal 720$^{\circ}$. The telescopes observe $\sim75\%$ of the sky every day. Complete measurement of Stokes Q and U is enabled by daily rotations of the telescope boresight platform, which rotate the orientation of the detector polarization on the sky. The rotation about the boresight spans $90^{\circ}$ as depicted in Fig.~\ref{fig:bs_rotation}, cycling through angles $-45^{\circ}$, $-30^{\circ}$, $-15^{\circ}$, $0^{\circ}$, $15^{\circ}$, $30^{\circ}$, $45^{\circ}$ each week. This scan strategy enables cross-linking on different time scales with each pixel on the sky being observed with different scanning directions and detector orientations.

\begin{figure*}[!htb]
    \centering
    \includegraphics[width=1\textwidth]{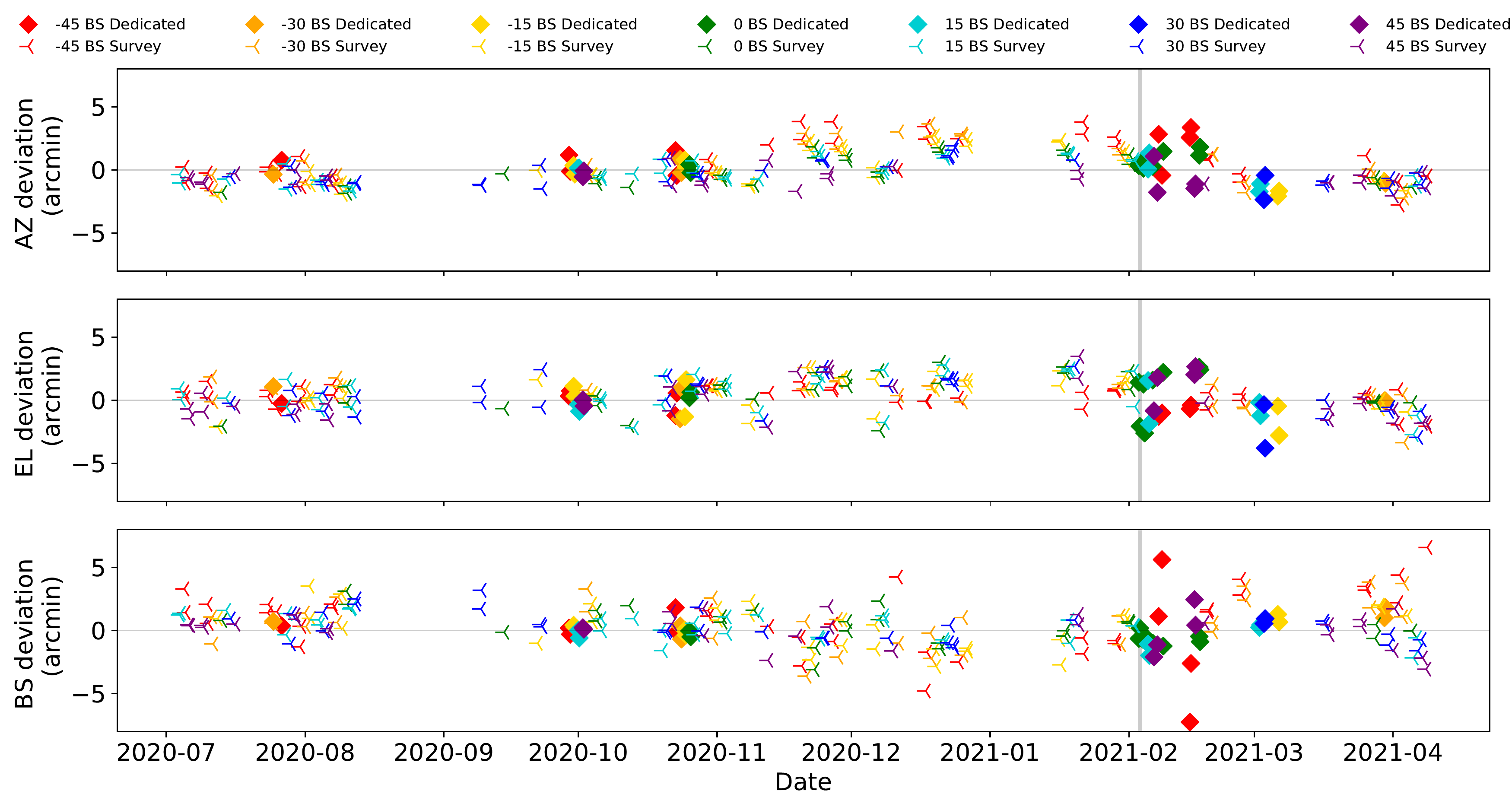}
    \caption{Deviations from the W-band pointing model in arcminutes for telescope azimuth, elevation, and boresight rotation angle over the period of July 2020 through April 2021. Different telescope boresight angles ranging from $-45^{\circ}$ to $+45^{\circ}$ are highlighted in a rainbow color pattern. Diamonds represent deviations inferred from {\textit{dedicated scans}}, and tri-left markers from {\textit{survey scans}}.  Only those {\textit{survey scans}} where at least 100 detectors see the Moon are used to derive pointing information. Each data point corresponds to pointing deviation inferred from all detectors that see the Moon during that {\textit{dedicated}} / {\textit{survey scan}} combined. Deviations are inferred separately from scans conducted during the Moon rising and setting, when available. The gray vertical line highlights the start date of a new pointing correction model based on a series of {\textit{dedicated scans}}.}
    \label{fig:W1_pointing}
\end{figure*}

\begin{figure*}[htb]
    \centering
    \includegraphics[width=0.9\textwidth]{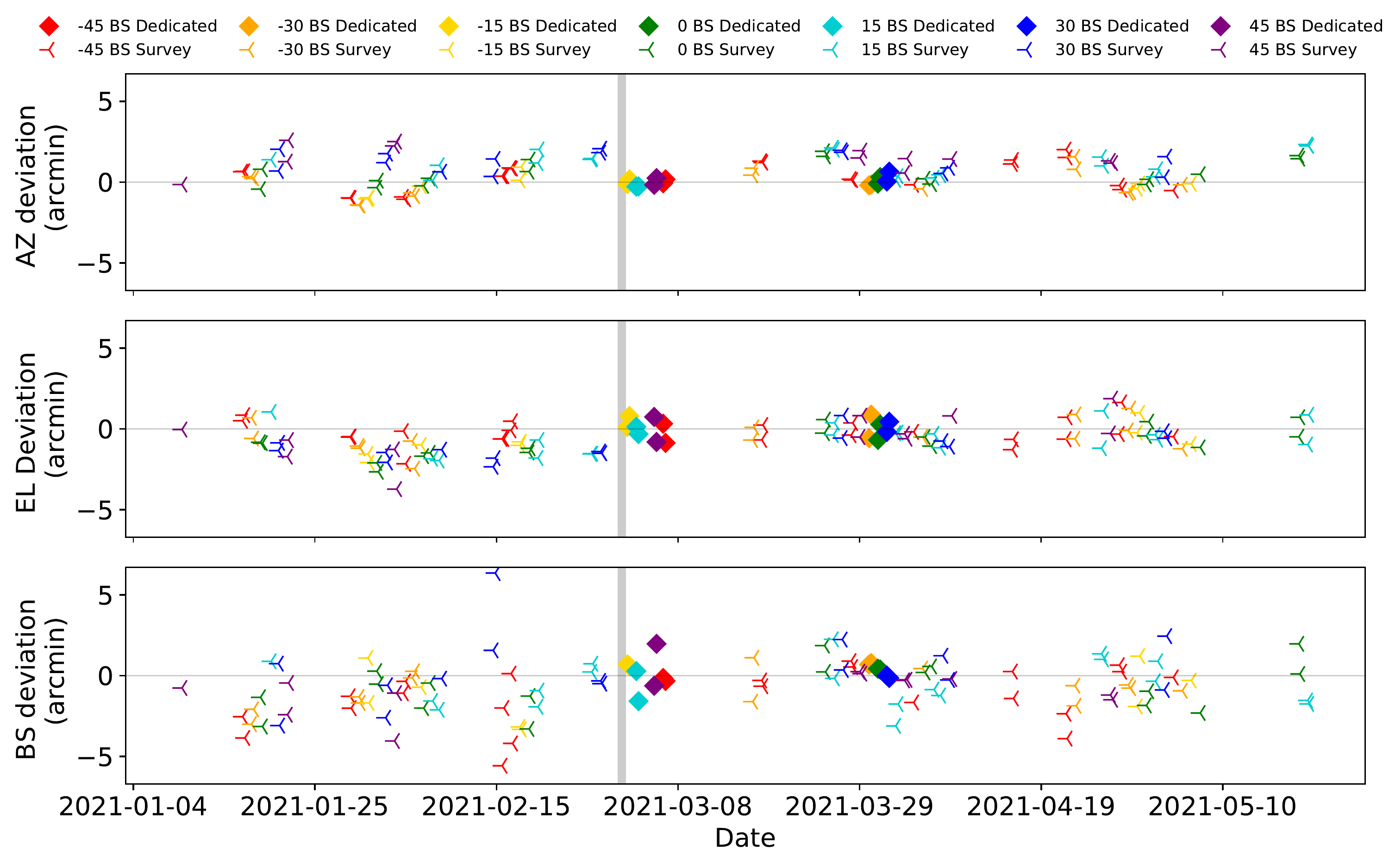}
    \caption{Deviations from the HF pointing model in arcminutes for telescope azimuth, elevation, and boresight rotation angle over the period of January through May 2021. Different telescope boresight angles ranging from $-45^{\circ}$ to $+45^{\circ}$ are highlighted in a rainbow color pattern. Diamonds represent deviations inferred from {\textit{dedicated scans}}, and tri-left markers from {\textit{survey scans}}. Only those {\textit{survey scans}} where at least 100 detectors see the Moon are used to derive pointing information. Each data point corresponds to pointing deviation inferred from all detectors that see the Moon during that {\textit{dedicated}} / {\textit{survey scan}} combined. Deviations are inferred separately from scans conducted during the Moon rising and setting, when available. The gray vertical line highlights the start date of a new pointing correction model based on a series of {\textit{dedicated scans}}.}
    \label{fig:HF_pointing}
\end{figure*}

Calibration campaigns are periodically undertaken, in which dedicated observations of bright sources, namely, the Moon, Jupiter, and Venus are performed. Moon scans are used for pointing calibration, whereas planet scans are used for beam characterization since the bright signal from the Moon saturates many of the detectors. The planets are essentially point sources given the instrument beam size at all frequencies. During each calibration run, the telescope scans across the source back and forth azimuthally, maintaining a constant elevation and allowing the source to drift through the FOV in elevation. For the remainder of the paper, a ``scan'' in the context of pointing calibration and beam characterization refers to a set of continuous sweeps conducted across a source as it rises or sets through the array. Scans performed with different boresight rotations help improve the sampling of the spatial beam and provide information about the offset of the center of the array which would otherwise be degenerate with the boresight pointing. In addition to these {\textit{dedicated scans}}, the telescopes scan over the Moon during CMB observations (henceforth referred to as {\textit{survey scans}}). The {\textit{dedicated scans}} are used for updating the pointing model while the {\textit{survey scans}} are only used for tracking pointing deviations from the last updated model, because the Moon sampling is sparse in such scans.

\section{POINTING}
\label{sec:pointing}  


The raw time-ordered data from the {\textit{dedicated scans}} are calibrated~\cite{2022arXiv220506901A}, filtered, and downsampled. Moon-centered intensity maps are then generated for each individual detector, which are fitted with two-dimensional Gaussian profiles to provide the pointing model. Details of the data processing will be described in a future paper. The pointing model is updated using the data from each calibration run. The pointing deviations with respect to the model in telescope azimuth, elevation, and boresight angle as a function of time over a representative period of several months during the 2020--2021 observing season are shown in Figures~\ref{fig:W1_pointing}~and~\ref{fig:HF_pointing} for 90~GHz and HF, respectively. The vertical gray lines indicate updates to the pointing model. The mean offset from model and standard deviations of the 90~GHz telescope pointing over this period of several months are $0.1'\pm1.4'$, $0.4'\pm1.3'$, and $0.4'\pm1.7'$ in azimuth, elevation, and boresight rotation angle, respectively. Similarly, the mean offset from model and standard deviations of the HF telescope pointing are $0.6'\pm0.9'$, $-0.5'\pm1.0'$, and $-0.6'\pm1.6'$ in azimuth, elevation, and boresight rotation angle, respectively. 


\section{BEAMS}
\label{sec:beams}  

{\textit{Dedicated scans}} of Jupiter performed around the time of the Earth's closest approach to Jupiter and during good weather conditions at the telescope site are used to measure the angular response of the detectors on the sky. Each {\textit{dedicated scan}} spans approximately two hours and consists of multiple passes over the planet. The detectors are tuned prior to each scan. For each detector and its corresponding pointing offset, we define a coordinate system centered at the detector, with the y-axis pointing along the local meridian in the receiver coordinate system, which is a spherical coordinate system with array center located at zero longitude on the equator. Using the data from each scan, a per-detector planet map is generated out to a radius of 4$^{\circ}$ and evaluated on criteria such as signal-to-noise ratio ($S/N$) and root-mean-square ($rms$) noise. Only maps meeting certain quality thresholds are retained. Details regarding these selection criteria will be described in a future paper. The retained maps are averaged per detector with $S/N$-weighting to generate average per-detector beam maps. A total of 319, 389, and 211 detectors at 90, 150, and 220~GHz, respectively, detected the signal from Jupiter~\cite{2022ApJ...926...33D}. For each detector, we define an effective beam full width at half maximum (FWHM) as
\begin{equation}
    \sigma_\mathrm{eff} = \sqrt{2\frac{\sigma_\mathrm{maj}^{2}\times\sigma_\mathrm{min}^{2}}{\sigma_\mathrm{maj}^{2}+\sigma_\mathrm{min}^{2}}}
    \label{equ:effective_fwhm}
\end{equation}
where the FWHM of the beams along the major and minor axes $\sigma_\mathrm{maj}$ and $\sigma_\mathrm{min}$ are obtained from a two-dimensional Gaussian fit to the average per-detector beam maps.

\begin{figure*}
\includegraphics[width=0.95\linewidth]{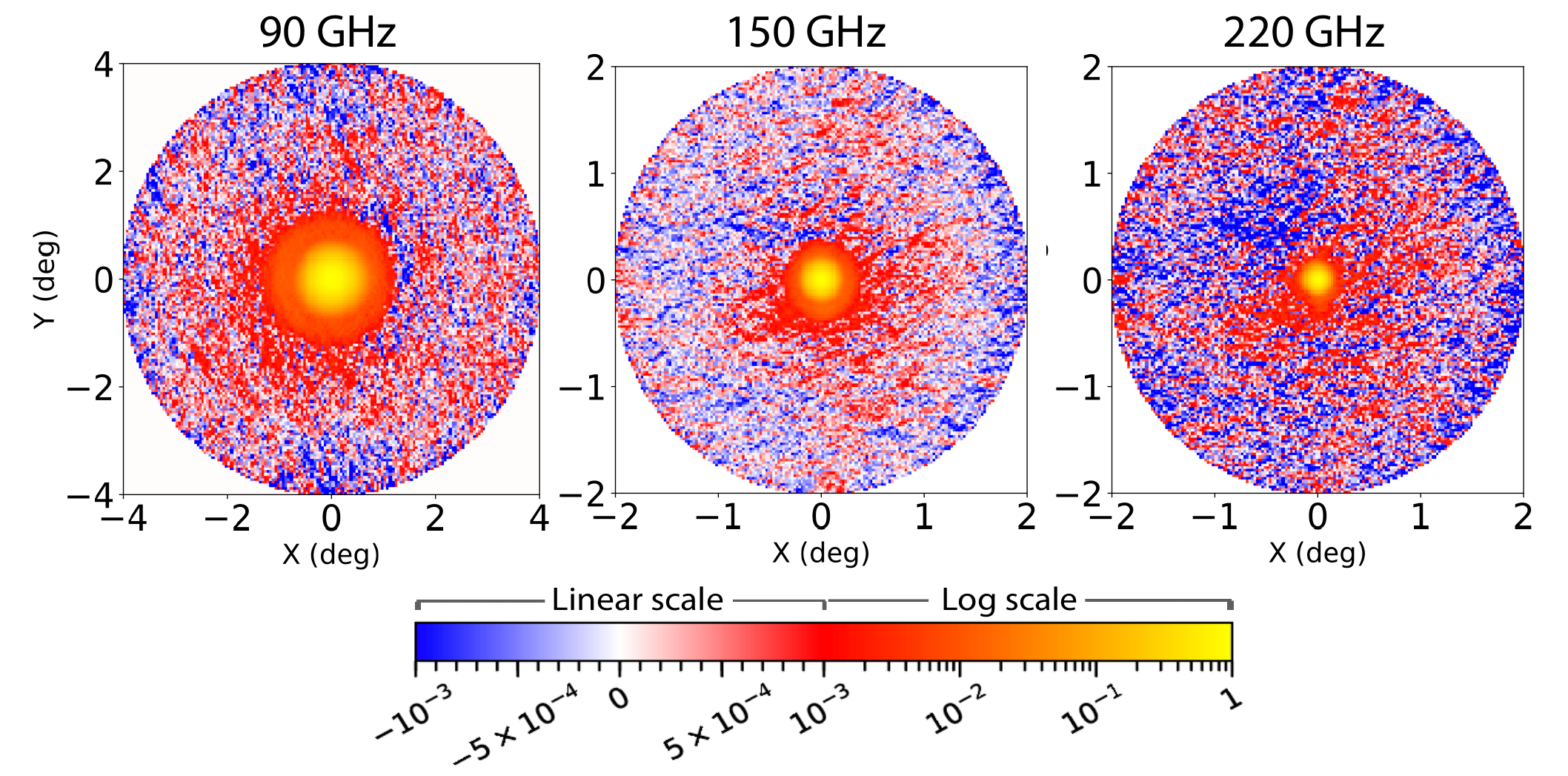}
\caption{\textbf{Average instrument beams}. Preliminary peak-normalized beam maps out to 4$^{\circ}$ in radius at 90~GHz (left panel) and 2$^{\circ}$ in radius at 150 and 220~GHz (middle and right panels, respectively) obtained by coadding 209, 385, and 208  average per-detector beam maps at 90, 150, and 220~GHz, respectively. The color bar scale is logarithmic from 1 to $10^{-3}$, and linear from $10^{-3}$ to $-10^{-3}$ in order to display negative values. The map has a resolution of 0.05$^{\circ}$. }
\label{fig:avg_beam_map}
\end{figure*}

\begin{figure}
\begin{center}
\includegraphics[width=0.65\linewidth]{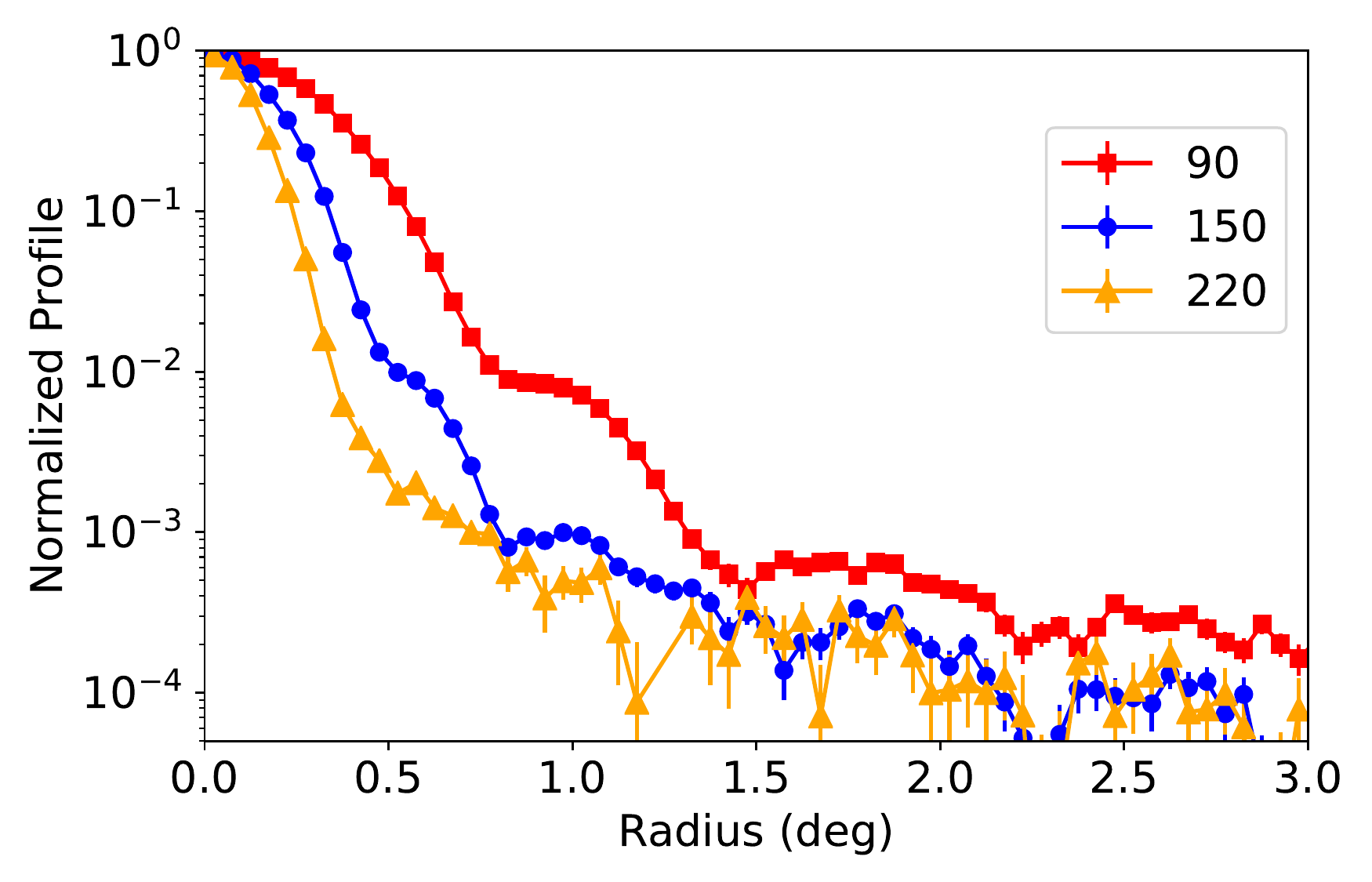}
\caption{\textbf{Beam profiles}. Azimuthally averaged 90 (red squares), 150 (blue circles), and 220 (orange triangles) GHz preliminary beam profiles out to 3$^{\circ}$. The errorbars represent 1$\sigma$ uncertainties.}
\label{fig:1D_beam_prof}
\end{center}
\end{figure}

The per-detector averaged maps are then $S/N$-weighted and co-added to generate the effective instrument beam map. When co-adding, we excluded detectors with (1) low end-to-end optical efficiency, (2) highly eccentric beams, or (3) an out-of-range voltage bias for the majority of planet scans.
In total, the above selection criteria resulted in 209, 385, and 208 per-detector beam maps being averaged to generate the effective instrument beam maps at 90, 150, and 220~GHz, respectively, as shown in Fig.~\ref{fig:avg_beam_map}. Due to the left-right symmetry of the optics, averaging per-detector maps from across the focal plane circularizes the instrument beam map. Additionally, boresight rotation and observations of the same point rising and setting (i.e., sky rotation) further ensure that the effective beam for CMB observations is nearly circular. Data from the seven boresight orientations contribute with nearly equal weights to the survey map. Therefore, we approximate our effective beams by azimuthally averaged radial beam profiles as shown in Figure~\ref{fig:1D_beam_prof}. 

\section{Summary}
\label{sec:summary}

In these proceedings, we presented measurements of the 90~GHz and HF telescope pointing stability over several months during the 2020--2021 observing season. The telescope pointing offset from model and pointing uncertainties in azimuth, elevation, and boresight rotation angle are on the order of a few percent of the respective beam FWHM. We also presented preliminary measurements of the 90, 150, and 220~GHz average instrument beam in intensity using dedicated planet observations. The measured effective beam FWHMs are $0.615\pm0.019^{\circ}$, $0.378\pm0.005^{\circ}$, and $0.266\pm0.008^{\circ}$ at 90, 150, and 220~GHz, respectively. Other studies related to pointing, beams and optical characterization will be discussed in a future paper.

\acknowledgments 
 
We acknowledge the National Science Foundation Division of Astronomical Sciences for their support of CLASS under Grant Numbers 0959349, 1429236, 1636634, 1654494, 2034400, and 2109311. We thank Johns Hopkins University President R. Daniels and the Deans of the Kreiger School of Arts and Sciences for their steadfast support of CLASS. We further acknowledge the very generous support of Jim and Heather Murren (JHU A\&S ’88), Matthew Polk (JHU A\&S Physics BS ’71), David Nicholson, and Michael Bloomberg (JHU Engineering ’64). The CLASS project employs detector technology developed in collaboration between JHU and Goddard Space Flight Center under several previous and ongoing NASA grants. Detector development work at JHU was funded by NASA cooperative agreement 80NSSC19M0005. CLASS is located in the Parque Astron\'omico Atacama in northern Chile under the auspices of the Agencia Nacional de Investigaci\'on y Desarrollo (ANID). We acknowledge scientific and engineering contributions from Max Abitbol, Fletcher Boone, Jay Chervenak, Lance Corbett, David Carcamo, Mauricio D\'iaz, Ted Grunberg, Saianeesh Haridas, Connor Henley, Ben Keller, Lindsay Lowry, Nick Mehrle, Grace Mumby, Diva Parekh, Isu Ravi, Daniel Swartz, Bingjie Wang, Qinan Wang, Emily Wagner, Tiffany Wei, Zi\'ang Yan, Lingzhen Zeng, and Zhuo Zhang. For essential logistical support, we thank Jill Hanson, William Deysher, Miguel Angel D\'iaz, Mar\'ia Jos\'e Amaral, and Chantal Boisvert. We acknowledge productive collaboration with Dean Carpenter and the JHU Physical Sciences Machine Shop team.
Sumit Dahal is supported by an appointment to the NASA Postdoctoral Program at the NASA Goddard Space Flight Center, administered by Oak Ridge Associated Universities under contract with NASA. Zhilei Xu is supported by the Gordon and Betty Moore Foundation through grant GBMF5215 to the Massachusetts Institute of Technology.

\bibliography{report} 
\bibliographystyle{spiebib} 

\end{document}